\documentclass[a4paper,preprint,aps,pre,showkeys,superscriptaddress,nofootinbib]{revtex4-1}    % aip substyle  jour. appl. phy. journal
\usepackage{graphicx}
\usepackage{amsmath}
\usepackage[usenames]{color}
\usepackage{dcolumn}%
\usepackage{bm}%

\newcommand{\bea}{\begin{eqnarray}}
\newcommand{\eea}{\end{eqnarray}}
\begin{document}

\title{Connectivity percolation in suspensions of hard platelets}

\author{Maneesh Mathew}
\email{mathewm@uni-mainz.de}
\affiliation{Institut f\"ur Physik, Johannes-Gutenberg-Universit\"at, D-55099 Mainz, Staudinger Weg 7, Germany}
\author{Tanja Schilling}
\affiliation{Theory of Soft Condensed Matter, Universit\'e du Luxembourg, 162a avenue de la fa\"iencerie, 1511 Luxembourg, Luxembourg}
\author{Martin Oettel}
\affiliation{Institut f\"ur Physik, Johannes-Gutenberg-Universit\"at, D-55099 Mainz, Staudinger Weg 7, Germany}
\affiliation{Institut f\"ur Theoretische Physik II, Heinrich-Heine-Universit\"at D\"usseldorf, Universit\"atsstr.~1, D-40225 D\"usseldorf, Germany}
\date{April 13, 2012}
\begin{abstract}
We present a study on connectivity percolation in suspensions of hard 
platelets by means of Monte Carlo simulation. We interpret our results using 
a contact-volume argument based on an effective single--particle cell model. 
It is commonly assumed that the percolation threshold 
of anisotropic objects scales as their inverse aspect ratio. While this rule 
has been shown to hold for rod-like particles, we find that for hard plate-like 
particles the percolation threshold is non-monotonic in the aspect ratio. 
It exhibits a shallow minimum at intermediate aspect ratios and then 
saturates to a constant value. 
This effect is caused by the isotropic-nematic transition pre-empting 
the percolation transition. Hence the common strategy to use highly 
anisotropic, conductive  particles as fillers in composite materials in order 
to produce conduction at low filler concentration is expected to fail for 
plate-like fillers such as graphene and graphite nanoplatelets.
\end{abstract}

\maketitle

\section{Introduction}
Composite materials of enhanced mechanical strength, thermal conductivity 
or electrical conductivity can be produced by mixing polymer resins with 
nanoparticles.
In particular carbon-based particles such as carbon nanotubes, graphene and 
graphite nanoplatelets are promising fillers because of their 
extraordinary materials properties \cite{moniruzzaman06, stankovich2006graphene, li2011}.
In this context percolation, a topic that had originally
been brought up in the context of fluid flow through 
porous media, has attained a new field of application.

The term percolation (more specifically connectivity percolation) refers to the 
transition at which a system 
changes from containing 
isolated clusters of particles to containing a 
system-spanning network that produces connectivity on a macroscopic 
scale \cite{stauffer-1994-book,meester-roy-book}.  
The density at which this 
transition occurs is called percolation threshold. 
A composite material made from an insulating matrix that is filled with 
conductive particles becomes conductive itself when the filler percolates 
\footnote{In contrast to the case of fluid flow through percolating cavities,
however, the filler particles do not need to touch each other. Their 
distances only need to be small enough to allow for tunneling of a 
sufficiently large number of charges.}.

For technological applications it is often desirable to use as low a 
filler content as possible (e.g.~because the filler material might be 
expensive, or it might have a deteriorating influence on other 
properties of a composite such as its transparency). 
As it has been argued that the percolation threshold of anisotropic objects 
should scale as their inverse aspect ratio \cite{balberg1984excluded},
one commonly uses highly anisotropic fillers such as metal fibres or 
carbon nanotubes. For rod-like fillers this has been shown to be a 
successful approach (see e.g.~\cite{Deng2009,Leung1991}).

Recently plate-like fillers have moved into the focus, in particular 
due to the development of graphite nanoplatelet 
reinforced polymer 
composites \cite{li2011, potts-graphene-composites-plmr-2011}. While graphite 
nanoplatelets share some of the promising materials properties of carbon 
nanotubes, the production of graphite nanoplatelets is in general less 
expensive and energy intensive. However, the percolation 
thresholds that were reached are also not as low as for carbon nanotubes. 
The authors of a recent review \cite{li2011} remark that experiments do not 
show the expected drop of the percolation threshold with increasing aspect 
ratio. 

The topic of percolation in three dimensional (3d) suspensions of platelets 
has been addressed in theoretical and numerical investigations 
\cite{charlaix-pt_random_array_discs-jphya-1986,PhysRevE.79.041134,ambrosetti2008percolative,quintanilla2007asymmetry,quintanilla2001measurement,
quintanilla2000efficient, otten2011connectivity}.
However, while much attention has been given to fully penetrable platelets, 
impenetrable (or otherwise interacting) platelets have, to our knowledge, 
only been studied in two cases: Ambrosetti and co-workers did a 
simulation study on percolation in hard, 
oblate ellipsoids \cite{ambrosetti2008percolative}, and Otten and van der 
Schoot addressed the issue briefly in an article on connectedness percolation 
theory of polydisperse fillers \cite{otten2011connectivity}. 
We will refer to their results in more detail in comparison to our 
results.

In this paper, we present a Monte Carlo simulation study on connectivity 
percolation in a suspension of hard platelets in 3d. 
We discuss, in particular, the dependence of the percolation threshold on 
the aspect ratio of the platelets, and the interference between the 
isotropic-nematic transition and the percolation transition. We interpret 
the simulation data by means of a contact-volume argument based on an effective single--particle cell model.

\section{Simulation Details}
As model platelets we used cut spheres. These objects are obtained 
when a sphere of diameter $D$ is intersected
with two planes parallel to the equatorial plane at a distance 
$L/2$ such that the sphere's caps are sliced off \cite{veerman1992phase}.
The interaction potential between two platelets is infinite if the 
platelets overlap and zero otherwise. With this interaction potential the 
system is purely entropic, its configurational properties 
do not depend on the temperature $T$. 
We use $L$ as the unit of length and $k_BT$ as the unit of energy, 
where $k_B$ is Boltzmann's constant.

Percolation occurs when the platelets form a system spanning cluster. 
A cluster is defined as two or more platelets which are connected. 
We consider two platelets as connected if their surfaces approach 
closer than a given value $A$. As we are using periodic boundary conditions, 
we regard a cluster as percolating if its particles are connected to their 
own periodic images. In a quantitative 
comparison to an experiment on electrical 
conductivity, the distance $A$ would 
correspond to the tunneling distance. However, the value of $A$ does not have a 
qualitative effect on the results we present, as long as it is smaller 
than the particle thickness $L$.

Monte Carlo simulations were performed for fixed volume $V$ and number of 
particles $N$. The volume of one particle is given by 
\bea
 \label{eq:vp}
  V_p = \frac{\pi L}{4} \left( D^2 - \frac{L^2}{3}\right) \;, 
\eea
and the volume fraction is defined by $\eta = N V_p/V$.
A cuboid simulation box of dimensions $B \times B \times B$ 
with periodic boundary conditions was used. 
Simulations for systems with a connecting distance $A=0.2L$ involved 
$1.5\times 10^6$ trial moves per particle for equilibration and $3\times 10^6$ 
moves per particle for sampling. For the other values of $A$ equilibration 
and sampling were performed each with $0.5\times 10^6$ moves per particle.   

The nematic order parameter $S$ was used to monitor the overall alignment 
of the platelets \cite{degennes-prost-book}. 
If $\hat{\mathbf{u}}_i$ is a unit vector perpendicular to 
the surface of platelet $i$, then $S$ is given 
by the largest eigenvalue of the tensor 
\[
\mathbf{Q}=\frac{1}{2N} \sum_{i}^{N}(3\hat{\mathbf{u}}_i \hat{\mathbf{u}}_i - \mathbf{I}) \quad ,
\] where  $\mathbf{I}$ is the identity matrix.
In a fully aligned configuration 
$S$ is unity and in a perfectly isotropic configuration it is zero.

\section{Results and Discussion}

\begin{figure}[h]
\centering
 \includegraphics[angle=-90,scale=0.40]{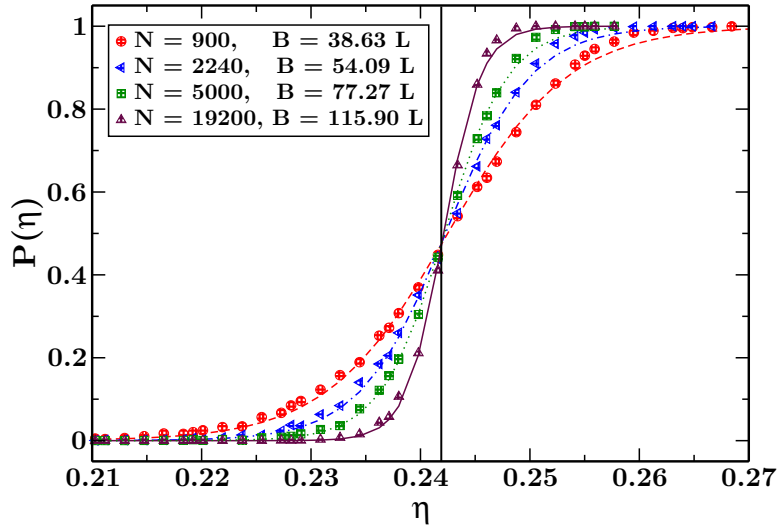}
\caption{Percolation probability as a function of volume fraction $\eta$ for different box sizes $B$ and $D/L=5$ ($N$ is the number of disks inside the box). 
The lines are fits by a tanh-profile, the solid vertical line marks the infinite system percolation threshold as obtained by a finite-size scaling analysis.}
\label{fig:finite-size-analysis}
\end{figure}

In an infinite system the percolation transition is marked by a 
discontinuity; the probability $P(\eta)$ to find a percolating cluster 
jumps from zero to one at the critical volume fraction  $\eta_c^{\rm inf}$. In 
a finite system, the transition is ``smeared out'', $P(\eta)$
forms a sigmoidal curve, which becomes steeper with increasing 
system size (see e.g.~ref.~\cite{stauffer-1994-book}).
In Fig.~\ref{fig:finite-size-analysis} we show the percolation probability 
for disks with aspect ratio $D/L=5$  as a function of volume fraction for 
several values of the box dimension $B$. The curves can be fitted by a 
profile of the form
\begin{equation}
 f=\frac{1}{2}\bigg\{1+\tanh\left[a(x-b)\right]\bigg\} \quad,
\label{eq:tanh}
\end{equation}
where $a$ is the inverse width of the transition region.
In order to extract $\eta_c^{\rm inf}$ from the simulation data we employ the 
finite-size scaling relation \cite{stauffer-1994-book, rintoul1997precise}. 
\begin{equation}
  a \propto B^{1/\nu} \quad ,
\end{equation}
where $\nu$ is the correlation length exponent. 
\begin{figure}[h]
\centering
 \includegraphics[angle=-90,scale=0.40]{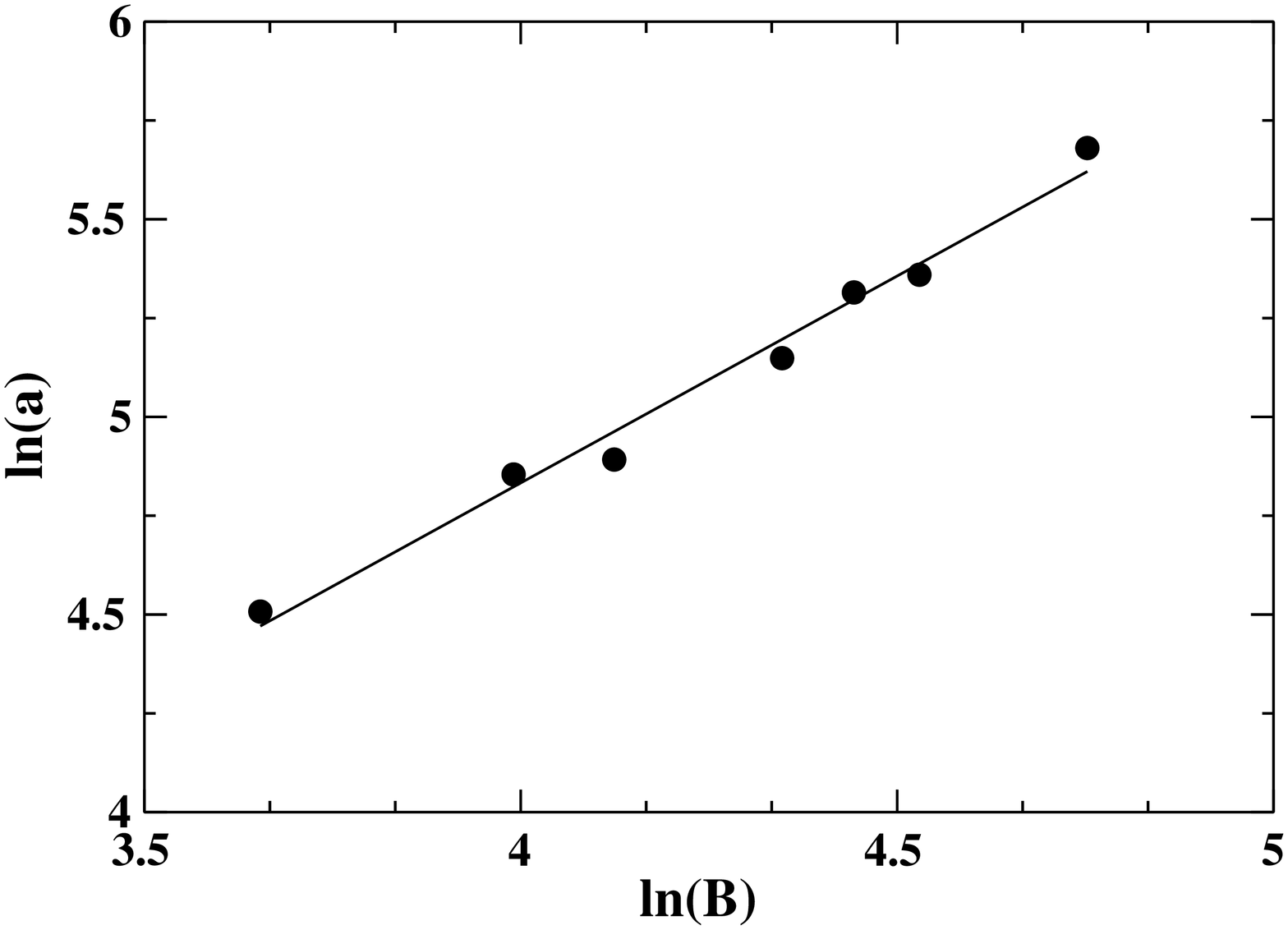}
\caption{Inverse transition width $a$ as a function of system size $B$ for $D/L=5$. The straight line is a fit to the data points. Its slope  
corresponds to a value of $\nu=0.96\pm0.01$.}
\label{fig:log-log}
\end{figure}
Fig.~\ref{fig:log-log} shows the inverse width $a$ for various system sizes. 
From the fitted line we find $\nu=0.96\pm0.01$, which is in agreement with the value 
for percolation in three dimensions (as well as with previous results on spheres and oblate 
ellipsoids \cite{rintoul1997precise,ambrosetti2008percolative}). 
\begin{figure}[h]
\centering
 \includegraphics[angle=-90,scale=0.40]{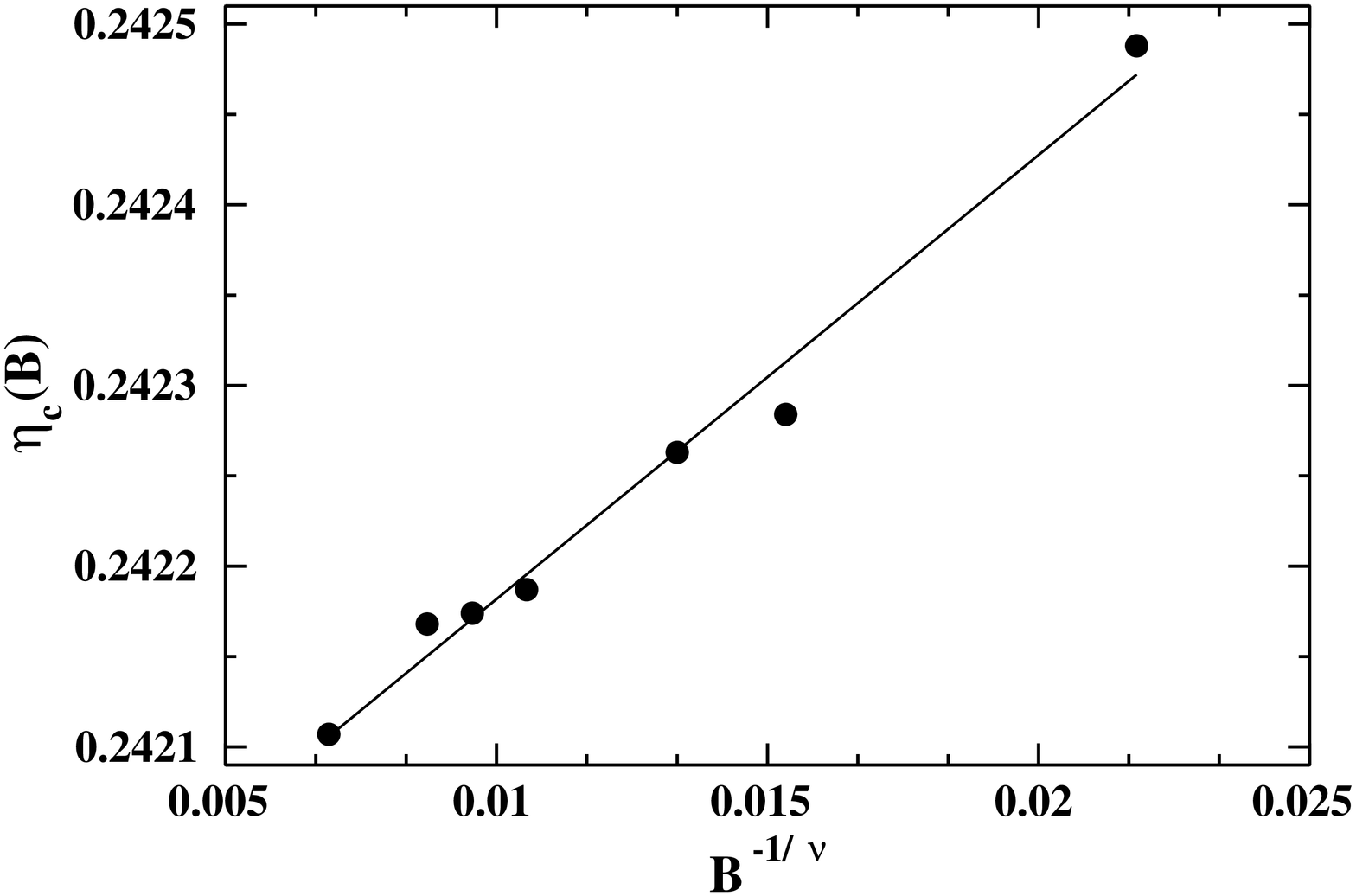}
\caption{Percolation threshold as a function of $B^{-1/\nu}$. The line is a fit to the data which yields $\eta_c^{\rm inf} = 0.2419\pm0.0001$.}
\label{fig:eta-b}
\end{figure}

The volume fractions at the inflection points of the tanh-profiles 
(eq.~\ref{eq:tanh}) 
can be interpreted as finite-size percolation thresholds $\eta_c(B)$. 
Fig.~\ref{fig:eta-b} shows $\eta_c(B)$ plotted 
versus $B^{-1/\nu}$. We fitted a line to the data using the value of $\nu$ 
obtained above and extrapolated the fit 
to $\eta_c^{\rm inf} = 0.2419\pm0.0001$. As the difference between $\eta_c(B)$ and $\eta_c^{\rm inf}$ is of order 
$10^{-3}$, and as we are interested in the qualitative behaviour of the percolation threshold with the 
aspect ratio, we used the volume fraction at the inflection point as an estimate for $\eta_c^{\rm inf}$ 
for all other simulation data presented in the following. 
(Often, instead of doing a finite-size scaling analysis on the locations of the inflection points, 
one uses the intersection points of the percolation curves as an estimate for the percolation threshold. 
In Fig.~\ref{fig:finite-size-analysis} we marked $\eta_c^{\rm inf} = 0.2419$ by a vertical line. 
$\eta_c^{\rm inf}$ coincides within $10^{-4}$ with the volume fractions of the intersection region.) 
 
\begin{figure}[t]
\centering
 \includegraphics[angle=-90,scale=0.40]{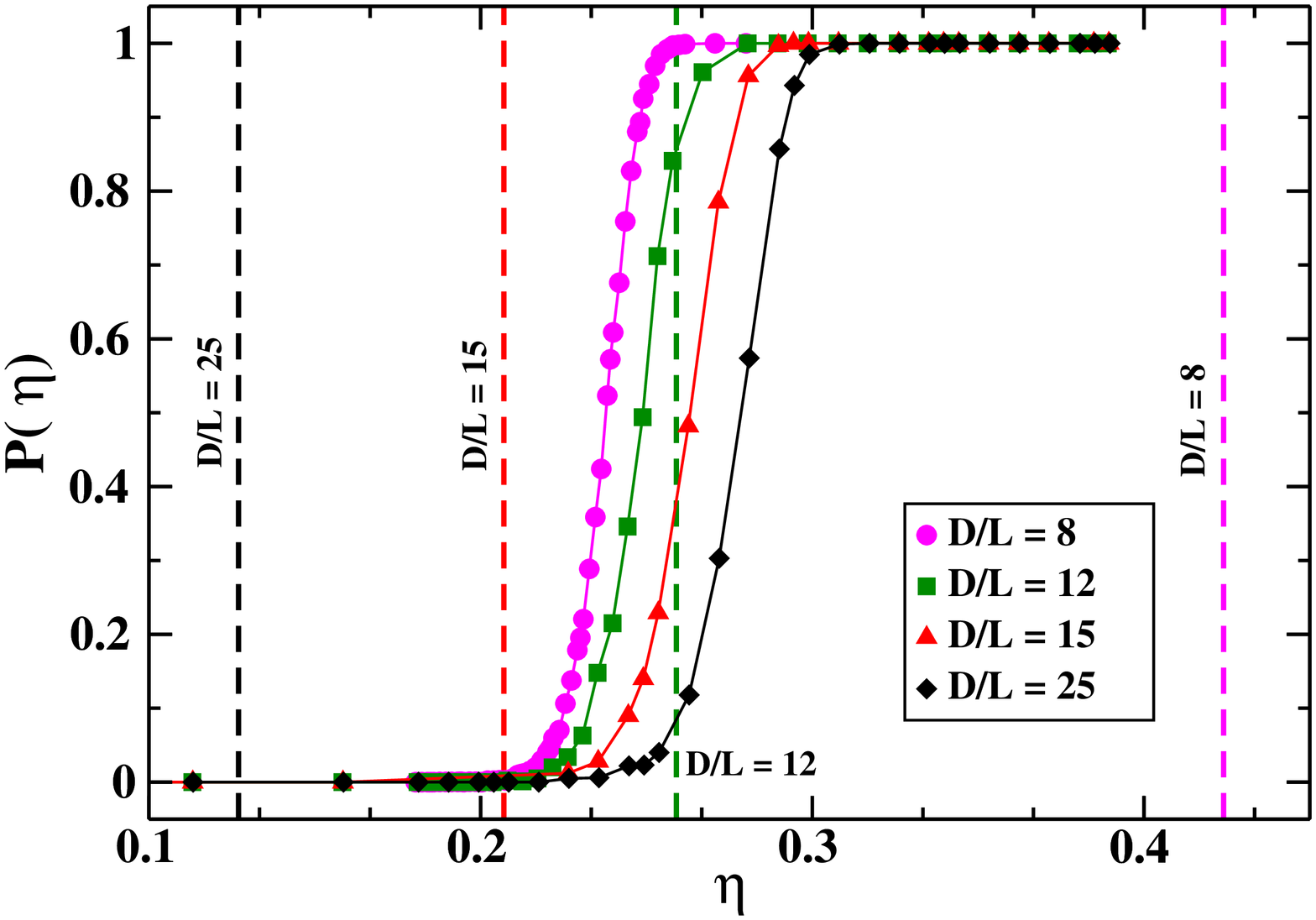}
\caption{Percolation probability as a function of volume fraction for different aspect ratios ($A=0.2 L$). 
The symbols and solid lines show the percolation probability, the dashed vertical lines indicate the volume fractions were the nematic order 
parameter $S=0.5$ (approximate location of the isotropic-nematic transition).}
\label{fig:perc-in-interference}
\end{figure}

In contrast to the rod-like systems which have been studied in 
the past, platelets exhibit an
interference of the percolation transition with the isotropic-nematic (IN)
transition, which leads to non-trivial behavior at the percolation threshold.
In Fig.~\ref{fig:perc-in-interference}, we show the percolation probabilty $P(\eta)$  as a function of
the volume fraction $\eta$ for various aspect ratios $D/L$ (connectivity distance $A/L=0.2$). The volume fractions at the corresponding IN transitions are 
marked by dashed lines. 
For smaller aspect ratios ($D/L \alt 8$), percolation occurs at much smaller
volume fractions than nematic ordering. For $D/L=12$ the onset of nematic 
order and the onset of percolation almost coincide.
For larger $D/L$ the situation is reverse: nematic ordering occurs now at 
much smaller volume fractions than percolation.  

Our central result is shown in Fig.~\ref{fig:pt-d}, which depicts 
the percolation threshold, $\eta_c$, as a function of the aspect ratio
$D/L$ for various connectivity distances $A/L = 0.2 \dots 1$. 
For a given $A/L$, $\eta_c$ begins at the known value for hard spheres ($D/L=1$) \cite{bug1985interactions,desimone1986theory}, then quickly rises to a maximum
at $D/L \sim 3$ and upon further increasing $D/L$  
falls off weakly -- until it reaches a minimum at intermediate aspect ratios. Upon further
increasing $D/L$, $\eta_c$ moderately rises again before finally reaching a plateau value. The effect of increasing
the connectivity distance consists in shifting the curves $\eta_c(D/L)$ to lower values and shifting the location
of the minimum to higher values of $D/L$. The occurence of the mininum 
and the subsequent plateau--like behavior for higher $D/L$ 
are consequences of the isotropic-nematic (IN) transition pre-empting 
the percolation transition. This can be seen e.g. when comparing 
Figs.~\ref{fig:perc-in-interference} and \ref{fig:pt-d} for $D/L=12$ 
and $A/L=0.2$. Thus when the behavior of $\eta_c(D/L)$ beyond the minimum is analyzed one 
should take into account the nematic order in the system. In order to develop 
an understanding, we discuss the different regimes in detail in turn. 

\subsection{The isotropic regime} 
\label{subsec:iso}
\begin{figure}[h]
\centering
 \includegraphics[angle=-90,scale=0.40]{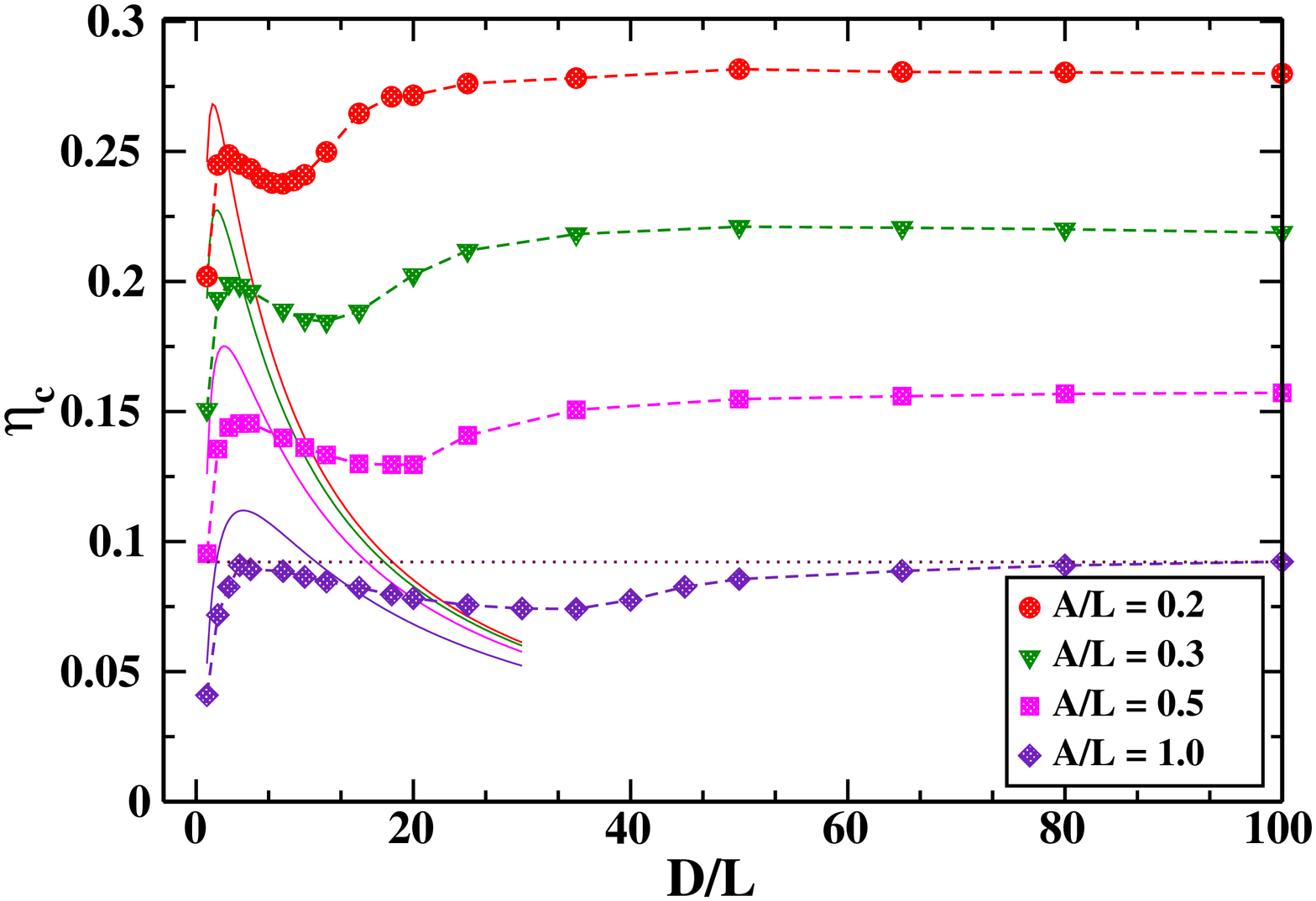}
\caption{Percolation threshold as a function of disk aspect ratio for various connectivity distances. 
The solid lines represent the percolation threshold according to the contact volume argument 
(Eq.~(\ref{eq:contactvolume}), $\alpha=1.7$). 
The dotted straight line is the constant value of $\eta_c$ predicted in Ref.~\cite{otten2011connectivity} 
for the case $A/L=1$.}
\label{fig:pt-d}
\end{figure}

Based on results for needle--like filler particles one would expect a 
decrease of $\eta_c$ with increasing 
aspect ratio \cite{Deng2009, Leung1991, schilling2007depletion}.
This kind of behavior is also predicted by a popular contact--volume argument which, in neglect of specific orientational correlations,
assumes that $\eta_c$ is proportional to the ratio
between the hard particle volume $V$ and the overlap volume $V_{\rm over}$ (second virial coefficient) 
of two hard particles \cite{balberg1984excluded}. 
The proportionality constant $\alpha$  can be loosely associated with the 
number of next neighbours (contacts) within
the overlap volume around one particle \cite{balberg1984excluded,schilling2007depletion}.

For hard disks we have
$V(D,L)=V_p$ (Eq.~(\ref{eq:vp})) and $V_{\rm over}(D,L)=(\pi D^3/3)(\cos\theta(1+\sin^2\theta)+3(\cos\theta+[\theta\sin\theta]/2)(\cos\theta+[\sin^2\theta]/2)$
with $\theta=\arccos (L/D)$ \cite{Fre89}. The connectivity criterion set by the minimal distance $A$ between two particles
can be incorporated by taking $V_{\rm over}(D+A,L+A)$ such that
\bea
 \label{eq:contactvolume}
 \eta_c \approx 2\alpha\; \frac{V(D,L)}{V_{\rm over}(D+A,L+A)} \quad . 
\eea
For $A \alt L \ll D$, the contact--volume argument predicts $\eta_c \propto L/D$, i.e. an inverse proportionality with 
the aspect ratio, similar to the case of rods. 

However, such a behavior is completely absent from the data. Our data
for the isotropic phase ($D/L$ below the minimum of $\eta_c(D/L)$) are not in 
the asymptotic regime in which Eq.~(\ref{eq:contactvolume}) could be applied 
(see Fig.~\ref{fig:pt-d}). Using the
full formula for the contact--volume argument (solid lines in 
Fig.~\ref{fig:pt-d}), the existence of a maximum 
near the sphere limit is predicted correctly,
however, the subsequent decay with $D/L$ is much too strong. 
Only for $A=L$ one sees a noticeably weaker
$D/L$--dependence. As a result we can state that neglecting orientational 
correlations leads to a prediction of the behavior of $\eta_c(D/L)$ that is  
qualitatively wrong. 
    
These findings can be checked against the Ornstein--Zernike approach to the connectedness 
correlation function. In a study on percolation of hard spheres \cite{desimone1986theory} 
the so-called ``bare chain sum approximation'' is introduced which amounts to the following overlap
volume criterion:
\bea
 \label{eq:conn_contactvolume}
 \eta_c \approx \alpha'\; \frac{V(D,L)}{V_{\rm over}(D+A,L+A)-V_{\rm over}(D,L)} . 
\eea
It suffers from an unphysical divergence of $\eta_c$ when $A\to 0$. For needle-like
particles (spherocylinders of length $L$, radius $R$, connectivity distance $2A$ with $L\gg R$) both
the contact volume argument (\ref{eq:contactvolume}) and the bare chain sum approximation
(\ref{eq:conn_contactvolume}) give the same dependence $\eta_c \propto 1/L$ but the volume
quotient in the bare chain sum approximation is larger by a factor $(1+A/R)/(A/R)$. 
For disks, however, the bare chain sum approximation ($D \gg L$) gives $\eta_c \propto L/A$ (independent of $D$)
instead of $\eta_c \propto L/D$ from the contact volume argument. This has been found also
in Ref.~\cite{ambrosetti2010solution} and is consistent with the recent work      
in Ref.~\cite{otten2011connectivity} using the
Ornstein--Zernike approach  
in the isotropic fluid. Here 
in the limit $A \approx L \ll D$  the same $D$--independent behavior is 
found: $\eta_{c,{\rm OZ}} \approx 2L/[A(5\pi+6)] $.
Indeed for $A=L$ this constant seems to be the correct first--order 
approximation (see dotted line in Fig.~~\ref{fig:pt-d}) but the other 
values for $A$ are apparently not compatible with the domain of validity  
($A\approx L$) for that result. 

The result of a percolation threshold that is independent of aspect ratio is 
in line with experimental observations \cite{li2011}. However, as the 
experimental samples are in general neither homogeneous nor in equilibrium, 
it is probably not the full explanation of this effect.

\subsection{The nematic regime}
\label{subsec:nematic}

\begin{figure}[th]
\centering
 \includegraphics[angle=-90,scale=0.29]{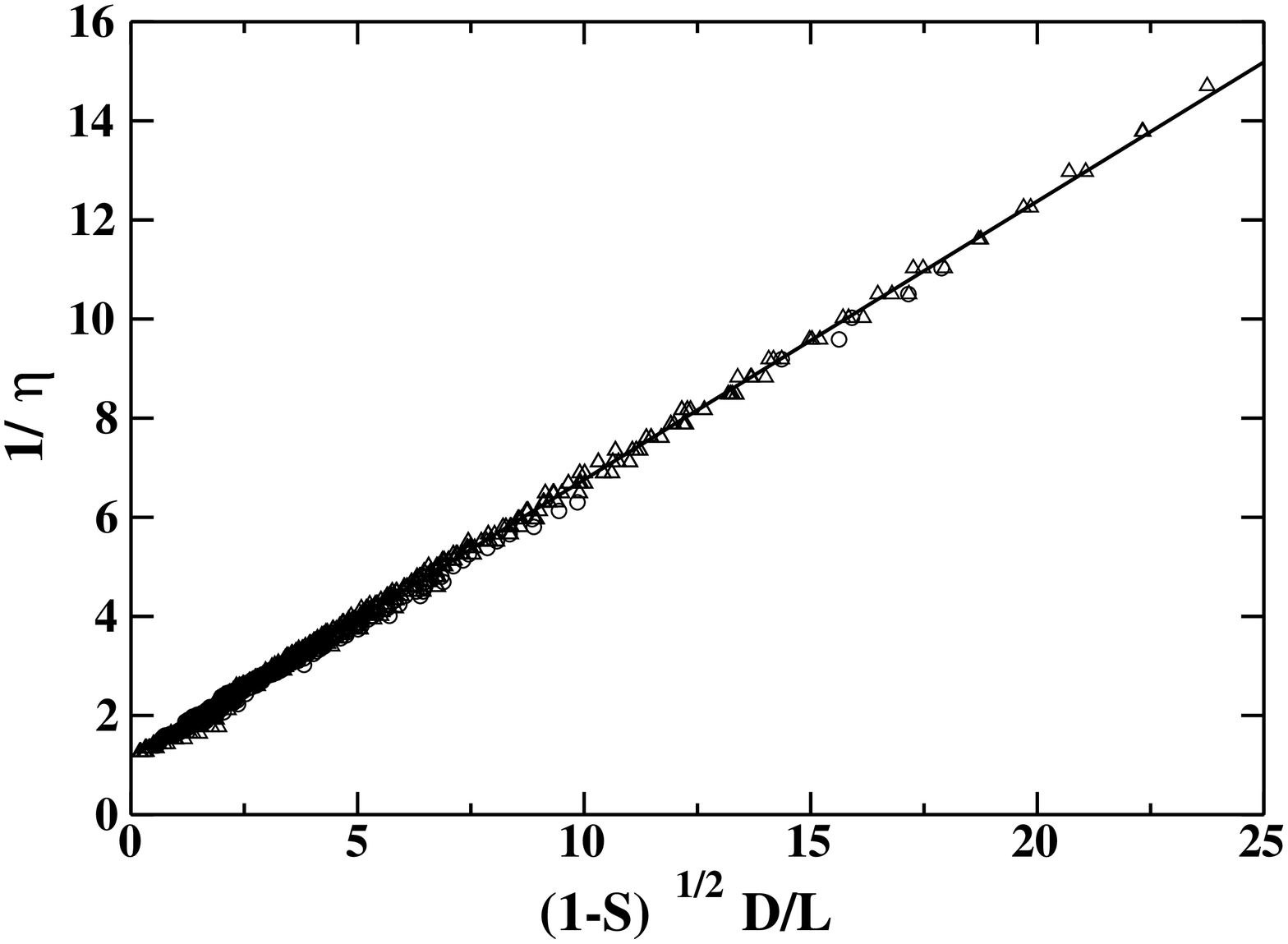}
 \includegraphics[angle=-90,scale=0.29]{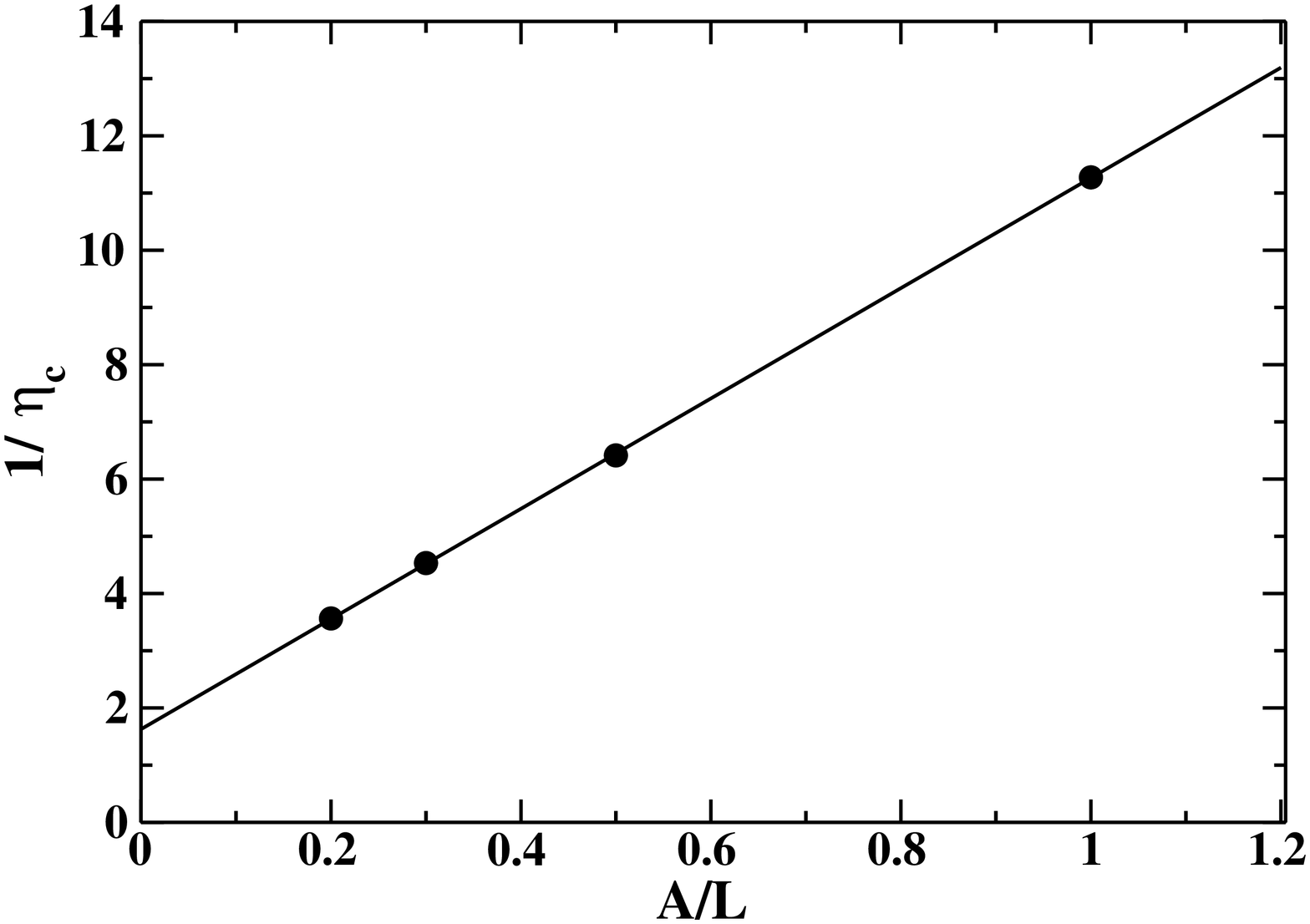}
\caption{ Left panel: Plot of inverse volume fraction $1/\eta$ vs. the scaling variable $\sqrt{1-S} D/L$ for our
data deep in the nematic phase with order parameter $S>0.9$. The linear fit assumes the cell theory equation
(\ref{eq:eosnem}). Right panel: Plot of inverse critical volume fraction $1/\eta_c$ at the plateau ($D/L \gg 1$)
 vs. connectivity parameter $A/L$. The linear fit is according to the volume argument in Eq.~(\ref{eq:volarg_nem}).}
\label{fig:eos_plateau}
\end{figure}

For high nematic order ($S>0.9$) the critical 
volume fraction $\eta_c(D/L)$ for percolation is in the plateau region 
(see Fig.~\ref{fig:pt-d}). For states with high order, one can invoke a single--particle cell model
to derive the following expression which connects volume 
fraction $\eta$, aspect ratio $D/L$ and order parameter $S$ 
\bea
 \label{eq:eosnem}
{\frac{\gamma}{\eta} =  1 + \beta  \frac{D}{L} \sqrt{1-S}} \quad .
\eea 
In the cell model, one assumes that each disk occupies an effective 
volume $V_{\rm eff}=(\pi/4)D^2 L_{\rm eff}$ and the effective height of the
disk is calculated via the one--particle orientation distribution function (see App.~\ref{app:eos}), 
$L_{\rm eff} = L + \beta D\sqrt{1-S}$. We determined the constant $\beta$ as 
0.49 by a fit to our data
(see Fig.~\ref{fig:eos_plateau}), being somewhat smaller than
derived from simple guesses for the orientation 
distribution ($\beta \sim 0.7$). The constant $\gamma=0.88$ is close to 1, 
which is consistent with the interpretation that the effective 
volume for all disks almost fills space.
A cell model description is known to be fairly accurate for the columnar phase of 
disks \cite{Wen04, Wen09} where one can go further and also
derive the full partition function. For the percolation problem, we are in 
the nematic phase but still far 
away from the columnar phase.  Hence it is
somewhat surprising that an effective one--particle argument  
quantifies  the value of the order parameter
for given volume fraction and aspect ratio also in the nematic phase.

In order to explain the plateau in Fig.~\ref{fig:pt-d}, we make the assumption 
that the onset of percolation is dictated by $A \propto L_{\rm eff} - L$. It 
is a similar argument as the contact volume argument in the isotropic phase,
namely that there is an invariant for percolation which is a quotient 
between two volumina: the effective
excess volume of one disk  $(\pi/4)D^2 (L_{\rm eff}-L)$ and the excess 
overlap volume of two {\em aligned} disks $\approx (\pi/4)D^2 A$.     
Using Eq.~(\ref{eq:eosnem}), we immediately find for the critical volume 
fraction at percolation:
\bea
 \label{eq:volarg_nem}
{\frac{\gamma}{\eta_c} =  1 + \beta'  \frac{A}{L}} \quad .
\eea 
With this assumption, $\eta_c$ is indeed independent of $D$ in the nematic 
phase, as seen in the data. In Fig.~\ref{fig:eos_plateau} 
we plot the plateau values of $1/\eta_c$ vs.~$A/L$ and also observe 
the predicted linear dependence. A linear fit, however,
gives $\gamma \approx 0.61$ (in contrast to $\gamma\approx 1$ from the 
equation of state (\ref{eq:eosnem}). 
The value of $\gamma$ is the critical volume fraction for
$A=0$ and thus corresponds to the volume fraction for the jammed state. 
This simple extrapolation, however, neglects the fact that the system undergoes
a transition to the columnar state before reaching such high values for the volume fraction.

\begin{figure}[h]
\centering
 \includegraphics[angle=-90,scale=0.40]{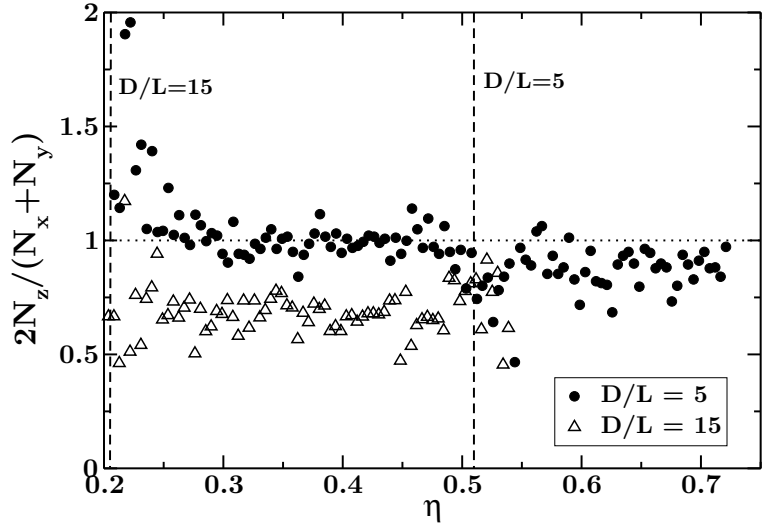}
\caption{Ratio of clusters that percolate along the director to clusters that percolate perpendicular to it. The vertical lines mark the IN 
transition ($S(\eta)=0.5$).}
\label{fig:percdir-d15}
\end{figure}

When the system is in the isotropic phase, the percolating cluster is 
connected to its periodic images in all spatial directions with equal 
probability. 
As the system undergoes the IN transition 
an asymmetry develops. In Fig.~\ref{fig:percdir-d15} we show the 
ratio of clusters that percolate ``along the director'' (in the $z$-direction) to
those that percolate ``perpendicular'' to the director, i.e. that are 
connected to their periodic images in the $x$- and/or $y$-direction. 
Once the system 
enters nematic phase, percolation along the director becomes less likely. 

\section{Summary and conclusions}

We have studied connectivity percolation for systems of monodisperse hard 
platelets modeled by cut spheres
of diameter $D$ and height $L$. The connectivity criterion between two 
particles was modeled by a minimum surface--to--surface distance $A$ chosen 
to be of the order of the particle height, $L \sim A$.  
For a fixed $A$, we have found that the critical 
volume fraction $\eta_c$ for percolation varies surprisingly little with the
aspect ratio $D/L$ (see Fig.~\ref{fig:pt-d}). The most notable 
feature of $\eta_c(D)$ is a 
shallow local minimum followed by a plateau for $D/L \gg 1$. We have 
demonstrated that the appearance of the
minimum is directly connected to the onset of nematic ordering in the 
system. The $D$--independent, plateau--like
behaviour of $\eta_c$ and the variation of the value of $\eta_c$ on the 
plateau with $A$ could be rationalized 
in terms of a contact--volume argument (see Sec.~\ref{subsec:nematic}). 
This argument relies on a derivation of the dependence of the nematic order parameter $S$ on aspect ratio $D/L$ 
and volume fraction $\eta$, based on an effective single--particle cell model.
We have tested this relation by our simulation results and found good 
agreement. In contrast to the nematic phase, a simple interpretation of 
$\eta_c(D/L)$ in the isotropic phase (small
aspect ratios) cannot be given, as angular correlations appear to play 
a major role. A theoretical
approach based on the solution of integral equations for the connectivity 
correlations appears to be promising
in this respect \cite{otten2011connectivity}, further work based on 
fundamental measure density functionals 
\cite{esztermann2006density,PhysRevE.81.041401} and related reference 
system closures \cite{Oet05,Aya09} is desirable in our opinion. 

It is a common strategy to use highly 
anisotropic, conductive  particles as fillers in 
composite materials to achieve conductivity at low filler concentrations. 
This strategy is based on a ``rule of thumb'' which says that $\eta_c$ 
decreases with the aspect ratio of anisotropic particles. A major conclusion 
from our work is that this rule does not hold for platelets such as graphite 
nano-platelets. This seems to be in contradiction to
the results in Ref.~\cite{ambrosetti2008percolative} where percolation 
of hard ellipsoids is studied and the corresponding $\eta_c$ for oblate 
ellipsoids (i.e.~disks) shows a strong decrease with aspect ratio
(see Fig.~3 in Ref.~\cite{ambrosetti2008percolative}). However, these data 
are obtained as curves $\eta_c(D/L)$ for constant $A/D$ (and not $A/L$ as 
in our case), i.e. along these curves the connectivity distance increases
with aspect ratio. We have re-analyzed the data in terms of a family of 
curves for constant $A/L$ and found that  apparently all data have been 
obtained in the isotropic phase and there is little variation of $\eta_c(D/L)$,
similar to our findings in the isotropic phase. Hence the authors of 
Ref.~\cite{ambrosetti2008percolative} could not see
the interference of the IN transition with the percolation transition.

\begin{acknowledgments}
The possibility to perform computer simulations at
the ZDV cluster in the Johannes Gutenberg Universit\"at Mainz is gratefully
acknowledged. M.~Mathew thanks the DFG for funding within the collabroative 
research center TR6, project D5 and the Center for Computational Sciences in Mainz (SRFN)
for financial support. We thank Claudio Grimaldi (EPFL Lausanne) and Paul van der Schoot (TU Eindhoven) for useful discussion.
\end{acknowledgments}

\begin{appendix}

\section{Cell model for strong nematic order}
\label{app:eos}

\begin{figure}
 \includegraphics[angle=0,scale=0.40]{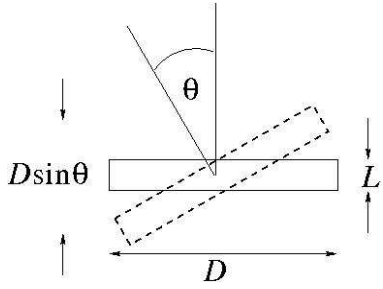}
 \caption{Geometrical definitions.}
 \label{fig:def}
\end{figure}

We consider disks deep in the nematic state, i.e.~$S \agt 0.9$, see 
Fig.~\ref{fig:def} for geometric definitions for one disk.
The deviation of the symmetry axis of one disk (disk orientation) with 
respect to the director is given by
$\theta$. The order parameter $S$ is defined by
\bea
 S = \bigg\langle \frac{3\cos^2\theta-1}{2} \bigg\rangle \;.
\eea
where $\langle \dots \rangle$ denotes the thermal average.
We call $p(\theta)$ the probability distribution function of disk 
orientations in the bulk state. Using this, the order parameter becomes
\bea
 S = \frac{1}{4\pi} \int d\Omega p(\theta) \frac{3\cos^2\theta-1}{2} = \frac{1}{2} \int_{-1}^1 d(\cos\theta) p(\theta)
   \frac{3\cos^2\theta-1}{2} \:.
\eea
Through fluctuations around the director, each disk will occupy approximately an effective volume ($D \gg L$):
\bea
 \label{eq:veff}
 V_{\rm eff} = \frac{\pi}{4} D^2 L_{\rm eff} = \frac{\pi}{4} D^2 \langle L + D\sin\theta \rangle \:.
\eea
The assumption of our cell model is that the whole system can be considered 
to be an ensemble of closely packed
effective volumes which are occupied by single disks which move independently 
of each other.
In order to determine that effective volume we consider a parametrization 
of the probability distribution function 
in the form $p(\theta) = N_\alpha p_{\rm red}(\theta/\alpha)$ where $\alpha$ 
is a width parameter in the angular variable
and $N_\alpha$ a normalization constant. Furthermore we consider narrow 
distributions (large $\alpha$) which permit us
to approximate angular averages  as 
\bea
   \frac{1}{2} \int_{-1}^1 d(\cos\theta) \dots p(\theta) \approx \int_0^\infty \theta d\theta \dots p(\theta)\:,
\eea
where the dots stand for any angular variable. Let us define moments of the 
reduced distribution function
by $M_n = \int_0^\infty x^n p_{\rm red}(x) dx$. Using these definitions, we 
have the following relations:
\bea
  1 = \int_0^\infty \theta d\theta p(\theta) = N_\alpha\; \alpha^2\; M_1 & \qquad& \mbox{(normalization)}\;, \\
  \frac{L_{\rm eff}-L}{D} = \int_0^\infty \theta^2 d\theta p(\theta) = N_\alpha\; \alpha^3\; M_2 & & \mbox{(effective height)}\;, \\
  1-S = \frac{3}{2} \int_0^\infty \theta^3 d\theta p(\theta) =\frac{3}{2} N_\alpha \; \alpha^4\; M_3 & & \mbox{(order parameter)}\;.
\eea
We can use the equations for normalization and order parameter to eliminate $\alpha$ and $N_\alpha$ and find
\bea
 \label{eq:leff}
 L_{\rm eff} = L + D \;\beta \sqrt{1-S} \\
             & & \left(\beta = \frac{M_2}{M_1}\sqrt{\frac{2}{3}\frac{M_3}{M_1}} \right) \nonumber  
\eea 
where $\beta$ depends only on the type of reduced distribution $p_{\rm red}$. As examples, for an exponential 
distribution $p_{\rm red}(x) = \exp(-x)$ (as found in the cell theory for the columnar phase \cite{Wen04,Wen09})
we have $\beta=2/3$ and for a Gaussian distribution $p_{\rm red}(x) = \exp(-x^2)$ we find
$\beta= \sqrt{\pi/6} \approx 0.72$.

As stated before, we proceed with the assumption that the sum of the 
effective volumes $V_{\rm eff}$
for $N$ particles (almost) fills the system volume $V$, 
i.e. $N V_{\rm eff}/V = \gamma \approx 1$.
Using the volume fraction $\eta = N (\pi/4) D^2 L /V$ ($D \gg L$) we 
rewrite this as
$\eta V_{\rm eff} / [(\pi/4) D^2 L] = \gamma$ and upon using Eqs.~(\ref{eq:veff}) and (\ref{eq:leff}) we
find the final result
\bea
 \gamma = \eta \left(  1 + \beta \sqrt{1-S} \frac{D}{L}  \right) \;.
\eea

\end{appendix}
% \bibliographystyle{plain}
%\bibliography{./plate_perc}

%

\end{document}